\documentclass[preprint,aps,prd,floatfix]{revtex4-2}
\usepackage{graphicx}
\usepackage{amsfonts}

\usepackage{amsmath}
\usepackage{rotating}
\usepackage{amssymb,amsthm}
\usepackage{soul}

\usepackage{xcolor}
\usepackage[normalem]{ulem}

\def \bh {\mbox{{\bf h}}}

\begin{document}


\title{Generalized Teleparallel de Sitter geometries}

\author{A. A. Coley}
\email{aac@mathstat.dal.ca}
\affiliation{Department of Mathematics and Statistics, Dalhousie University, Halifax, Nova Scotia, Canada, B3H 3J5}

\author {A. Landry}
\email{a.landry@dal.ca}
\affiliation{Department of Mathematics and Statistics, Dalhousie University, Halifax, Nova Scotia, Canada, B3H 3J5}

\author{R. J. van den Hoogen}
\email{rvandenh@stfx.ca}
\affiliation{Department of Mathematics and Statistics, St. Francis Xavier University, Antigonish, Nova Scotia, Canada, B2G 2W5}

\author{D. D. McNutt}
\email{david.d.mcnutt@uit.no}
\affiliation{Department of Mathematics and Statistics, UiT: The Arctic University of Norway, N-9037, Tromso, Norway}







\begin{abstract}

\vspace*{0.5cm}

Theories of gravity based on  teleparallel geometries are characterized by the torsion, which is a function of the coframe, derivatives of the coframe, and a zero curvature and metric compatible spin-connection. The appropriate notion of a symmetry in a teleparallel geometry is that of an affine symmetry. Due to the importance of the de Sitter geometry and Einstein spaces within General Relativity, we shall describe teleparallel de Sitter geometries and discuss their possible generalizations. In particular, we shall analyse a class of Einstein teleparallel geometries which have a $4$-dimensional Lie algebra of affine symmetries, and display two one-parameter families of explicit exact solutions.
%

%

\end{abstract}

\maketitle

\newpage


\section{Introduction}

Teleparallel geometries, which are characterized by the torsion (which is a function of the coframe and its derivatives), and a zero curvature and metric compatible spin-connection, 
provide an alternative to Riemmannian geometries in which to formulate a theory of gravity.  A particular subclass of teleparallel gravitational theories is dynamically equivalent to General Relativity (GR) and is often referred to as the Teleparallel Equivalent to General Relativity (TEGR) \cite{Aldrovandi_Pereira2013}. { In the {\it covariant} approach to the more general class of $F(T)$ teleparallel gravity theories  \cite{Krssak:2018ywd}, the teleparallel geometry is defined in a gauge invariant manner as a geometry having a spin connection with vanishing curvature.  The spin connection can be zero in a very special class of ``proper'' frames and is non-zero in all other frames \cite{Aldrovandi_Pereira2013,Krssak:2018ywd}.}

The torsion scalar, $T$, is defined in terms of the torsion tensor 
\begin{equation}
T=\frac{1}{2}T^a_{\phantom{a}\mu\nu}S_a^{\phantom{a}\mu\nu},
\end{equation}
where the superpotential, $S^a_{~\mu \nu}$, is constructed from the torsion tensor by
\begin{equation}
S_a^{\phantom{a}\mu\nu}=\frac{1}{2}\left(T_a^{\phantom{a}\mu\nu}+T^{\nu\mu}_{\phantom{\nu\mu}a}
    -T^{\mu\nu}_{\phantom{\mu\nu}a}\right)-h_a^{\phantom{a}\nu}T^{\phi\mu}_{~~\phi} + h_a^{\phantom{a}\mu}T^{\phi\nu}_{~~\phi}. \label{super}
\end{equation}

The variations of the Lagrangian $\frac{h}{2\kappa}F(T)+\mathcal{L}_{Matt}$, which include a non-trivial spin-connection \cite{Krssak:2018ywd,Krssak_Saridakis2015}, lead to {\it Lorentz covariant} { field equations (FEs)} and the resulting theory is locally Lorentz invariant \cite{Krssak_Pereira2015}. Treating the spin-connection as an independent field, the Lagrangian can then be written using Lagrange multipliers to impose the two constraints of zero curvature and  metric compatibility. Assuming an orthonormal frame, the corresponding variations { with respect to (w.r.t.)} the Lagrange multipliers lead to the following equations { (eqns.)}
\begin{equation}
\omega^a_{\phantom{a}b\mu} = \Lambda^a_{\phantom{a}c}\partial_\mu\Lambda_{b}^{\phantom{a}{c}} \quad \ \mbox{and}\ \quad \omega_{(ab)\mu} = 0 , \label{solution_omega}
\end{equation}
where  $\Lambda^a_{\phantom{a}b}\in SO(1,3)$  (and $\Lambda_b^{\phantom{b}c}
\equiv (\Lambda^{-1})^c_{\phantom{c}b}$).
Variations of the Lagrangian  with respect to the coframe can be decomposed into symmetric and antisymmetric parts:
\begin{subequations}
\begin{eqnarray}
\kappa \Theta_{(ab)} &=& F''(T)S_{(ab)}^{\phantom{(ab)}\nu} \partial_{\nu} T+F'(T){\tilde{G}}_{ab}  + \frac{1}{2}g_{ab}\left(F(T)-TF'(T)\right),\label{temp1}
        \\
             0      &=& F''(T)S_{[ab]}^{\phantom{[ab]}\nu} \partial_{\nu} T,\label{temp2}
\end{eqnarray}
\end{subequations}
where ${\tilde{G}}_{ab}$ is the Einstein tensor calculated explicitly from the metric \cite{Krssak:2018ywd}. Note that any quantity with an overtilde is computed using the Levi-Civita connection in terms of the metric. In the above FEs, $\Theta_{(ab)}$ is the matter energy momentum tensor (and $\Theta_{[ab]}=0$ due to the invariance of the FEs under $SO(1,3)$).

{

The determination of new solutions in a given $F(T)$ theory is vital for investigating the validity of the theory. Furthermore, distinguishing a new solution from a known solution can be difficult from the perspective of coordinates, as this necessitates finding the coordinate transformation that brings one solution to the other. As an alternative, it is possible to use torsion invariants to determine the equivalence of two solutions. The Cartan-Karlhede algorithm is an algorithmic approach to generate a geometrically preferred frame and connection which can characterize a given solution in terms of invariant quantities \cite{Coley:2019zld}.

The algorithm exploits the canonical form of the $q$-th order torsion tensors (that is the torsion tensor and its $q$-th covariant derivatives) to fix the frame. At the zeroth iteration, the torsion tensor is computed and the Lorentz frame freedom is used to adapt the frame so that the torsion tensor takes on a canonical form. The number of functionally independent components, $t_0$, are recorded and the remaining frame freedom, which does not affect the torsion tensor, is denoted as the $0$-th order linear isotropy group, $H_0$.

For the $q$-th iteration, the $q$-th order torsion tensor is computed and the $q-1$ linear isotropy group, $H_{q-1}$ is used to determine a canonical form for the $q$-th torsion tensor and the number of functionally independent components, $t_q$, are recorded along with the $q$-th linear isotropy group, $H_q$. The algorithm stops at iteration $p+1$, when $t_p = t_{p+1}$ and $H_p = H_{p+1}$. The subgroup $H_p$ of the Lorentz group is then called the linear isotropy group of the geometry.

The resulting invariant quantities that characterize the geometry are then $\{ t_q\}_{q=0}^{p+1}$, $\{ H_q \}_{q=0}^{p+1}  $ and the set of components $\mathcal{T} = \{ T^a_{~bc}, \nabla_{d_1} T^a_{~bc}, \ldots, \nabla_{d_{p+1}} \ldots \nabla_{d_1}  T^a_{~bc} \}$. The set $\mathcal{T}$ are torsion invariants known as {\it Cartan invariants}. If two solutions share the same discrete sequences, with Cartan invariants $\mathcal{T}$ and $\mathcal{T'}$, respectively,  equivalence of the solutions can be proven in a non-algorithmic way by solving the equations arising from equating the components of $\mathcal{T}$ and $\mathcal{T'}$. 

}

In teleparallel geometries, the tetrad (or (co)frame) and the corresponding spin-connection replace the metric as the primary objects of study. Consequently, the appropriate notion of symmetry in a teleparallel geometry is that of an affine symmetry (and not a metric symmetry; i.e, a Killing vector). An affine frame or intrinsic symmetry on the frame bundle is a diffeomorphism from the manifold to itself which is characterized by the existence of a vector field, ${\bf X}$, satisfying \cite{Coley:2019zld,McNutt_Coley_vdH2022}:
\begin{equation}
\mathcal{L}_{{\bf X}} \bh_a = \lambda_a^{~b} \,\bh_b \mbox{ and } \mathcal{L}_{{\bf X}} \omega^a_{~bc} = 0, \label{Intro:FS2}
\end{equation}
where $\omega^a_{~bc}$ denotes the spin-connection relative to the geometrically preferred invariantly defined frame $\bh_a$ determined by the Cartan-Karlhede algorithm and $\lambda_a^{~b}$ is an element of the linear isotropy group determined by the algorithm.

There do not exist any teleparallel geometries admitting a maximal group of affine frame symmetries other than Minkowski space \cite{Coley:2019zld,HJKP2018}. If a  four-dimensional teleparallel geometry has a non-zero torsion, then the maximum dimension of the group of affine symmetries is at most seven \cite{Coley:2019zld}. Therefore, in analogy with  de Sitter geometries in GR, we define the teleparallel de Sitter (TdS) geometry as that nontrivial teleparallel geometry which has a seven dimensional group of symmetries that is also a subgroup of the group of the Killing symmetries of the de Sitter metric \cite{preprint}.


Due to the importance of the de Sitter geometry within GR, we shall study TdS geometries and their generalizations in this paper. After a brief introduction to the TdS geometry, we discuss generalizations of this geometry mathematically and describe some physical applications, with particular emphasis on possible analogues of Einstein spaces. In particular, we shall discuss a class of Einstein teleparallel geometries which have a $4$-dimensional Lie algebra of affine symmetries (in the case where the Einstein space parameter $\lambda$ is equal to zero). We display the governing symmetric FEs explicitly in the first of the two solutions of the antisymmetric FEs. We shall investigate power law solutions and display two  one-parameter families of explicit solutions (where we can implicitly find the form of $F(T)$ for each parameter value).


\section{Teleparallel ``\lowercase{de} Sitter'' (T\lowercase{d}S)}

We { will} work  in a coframe in which the tangent space metric has the form $g_{ab}= \eta_{ab} = \mbox{Diag}[-1,1,1,1]$, which still allows a $O(1,3)$ subgroup of $GL(4,\mathbb{R})$ of residual gauge transformations which leaves the metric $g_{ab}=\eta_{ab}$ invariant.  We can then restrict attention to proper ortho-chronous Lorentz subgroups, $SO(1,3)$ or $SO(1,3)^+$.
Within this orthonormal gauge choice, the resulting field equations transform homogeneously under the remaining $O(1,3)$ (or $SO(1,3)$ or $SO(1,3)^+$)  Lorentz gauge transformations.

Using the Cartan-Karlhede algorithm \cite{Coley:2019zld}, we can determine the $G_7$ geometries obtained by requiring that the Cartan invariants are all constant. This follows from the formula for the dimension of the affine frame symmetry group,
$N = s+4-t_p$, where $s$ is the dimension of the linear isotropy group and $t_p$ is the number of functionally independent invariants at the conclusion of the algorithm. If all of the Cartan invariants are constant, then $t_p = 0$ and the dimension of the linear isotropy group is three, yielding a seven-dimensional affine frame symmetry group.

Solving the differential equations arising from the requirement that the only non-trivial components of the Cartan invariants are $T_{abc}$ yields the 
general case in which $a(t) = A_0 e^{H_0 t}$ is the scale factor, $k=0$ and $H_0$ is a non-zero constant in the teleparallel Robertson-Walker geometry \cite{newColey2023}. 
{ The special teleparallel geometry case where $H_0=0$, $a(t) = A_0$ a non-zero constant, $k=\pm 1$ and the additional affine frame symmetry $X_7 = \partial_t$ can be considered as the analogue of the Einstein static geometry in GR \cite{preprint} }. In the general case ($H_0\not =0$), the affine frame symmetry is of the form
\begin{equation}
X_7 = -\frac{1}{H_0} \partial_t + r \partial_r.  \label{X7_desitter}
\end{equation}
The resulting Lie algebra of ${ \{X_I\}_{I=1}^{7}} = \{X_1,X_2,X_3, X_4, X_5, X_6, X_7\}$ is given by
\begin{equation}
\begin{array}{lll}
{}  [X_1,X_5]=X_3,  & [X_1,X_6]=X_2,  & [X_1,X_7]=X_1, \\
{} [X_2,X_4]=-X_3, & [X_2,X_6]=-X_1, & [X_2,X_7]=X_2, \\
{}  [X_3,X_4]=X_2,  & [X_3,X_5]=-X_1, & [X_3,X_7]=X_3, \\
{}  [X_4,X_5]=-X_6, & [X_4,X_6]=X_5,  & [X_5,X_6]=-X_4.
\end{array} \nonumber
\end{equation}
This is a subalgebra of the Lie algebra for the group of metric (Killing) symmetries of de Sitter metric.
We therefore define
the Teleparallel de Sitter  (TdS) geometry as the teleparallel geometry with a $G_7$ Lie group of affine symmetries which is 
the semi direct product of the one-dimensional subgroup of $O(1,4)$ and the six dimensional Euclidean group $E(3)$.
Note in this geometry the covariant derivative of the torsion tensor is zero.


\subsection{Properties}
 
The diagonal de Sitter co-frame is
\begin{equation}
    h^a= \left[\begin{array}{c} dt\\
    e^{H_0t}\, dr \\
    e^{H_0t}\, rd\theta\\
    e^{H_0t}\, r\sin(\theta)d\phi
    \end{array} \right] , \label{deSitter_frame}
\end{equation}
which has the corresponding spin-connection one form
\begin{equation}
    \omega^a_{~b}= \left[\begin{array}{cccc}
    0 & 0& 0& 0 \\
    0 & 0 & -d\theta & -\sin(\theta) d\phi \\
    0 & d\theta & 0 & -\cos(\theta) d\phi \\
    0 & \sin(\theta) d\phi & \cos(\theta) d\phi & 0
    \end{array} \right]  , \label{desitter_conn}
\end{equation}
or the non-trivial components of the spin-connection are (all indices down)
\begin{eqnarray}
\omega_{23\theta}=-\omega_{32\theta} &=& -1 ,
\nonumber\\
\omega_{24\phi}=-\omega_{42\phi} &=& -\sin(\theta) ,
\nonumber\\
\omega_{34\phi}=-\omega_{43\phi} &=& -\cos(\theta)
\end{eqnarray}
(alternatively, one could use the proper de Sitter coframe given in \cite{preprint}). Using eqns. (\ref{deSitter_frame}) and (\ref{desitter_conn}) we have that
\begin{equation}
    T^a=H_0  \left[\begin{array}{c}
    0 \\
       h^1  \wedge h^2 \\
       h^1 \wedge h^3 \\
       h^1 \wedge h^4
    \end{array} \right] .
\end{equation}
The non-trivial components of the torsion tensor are
\begin{eqnarray}
T^2_{~tr}=-T^2_{~rt}            &=& H_0e^{H_0t} ,
\nonumber\\
T^3_{~t\theta}=-T^3_{~\theta t} &=& H_0e^{H_0t} \,r ,
\nonumber\\
T^4_{~t\phi}=-T^4_{~\phi t}     &=& H_0e^{H_0t} \,r\sin(\theta) .
\end{eqnarray}
{ The torsion tensor is decomposable into three irreducible parts \cite{Coley:2019zld}:
\begin{eqnarray}
T_{abc} = \frac{2}{3}\,\left(t_{abc}-t_{acb}\right) -\frac{1}{3}\,\left(g_{ab}\, V_c-g_{ac}\,V_b\right)+\epsilon_{abcd}\,A^d , \label{tensorparts}
\end{eqnarray}
where the vector, axial and tensor components are, respectively:
\begin{subequations}
\begin{eqnarray}
V_a &=& T^b_{\;\;ba} , \label{tensorpartvec}
\\
A^a &=& \frac{1}{6}\,\epsilon^{abcd}\,T_{bcd}, \label{tensorpartxi}
\\
t_{(ab)c} &=& \frac{1}{2}\,\left(T_{abc}+T_{bac}\right) -\frac{1}{6}\,\left(g_ {ca}\,V_b+g_{cb}\,V_a\right)+\frac{1}{3}\,g_{ab}\,V_c . \label{tensorpartsten}
\end{eqnarray}
\end{subequations}
} 
The dual superpotential two form ${}^*S_{a}$ is 
\begin{equation}
    {}^*S_a = \left[ \begin{array}{c}
    0 \\
    -2 H_0e^{2H_0t}\,    r^2\sin(\theta)d\theta \wedge d\phi \\
     2 H_0e^{2H_0t}\,    r\sin(\theta) dr \wedge d\phi \\
    -2 H_0e^{2H_0t}\,    r dr \wedge d\theta
    \end{array} \right] .
\end{equation}
The non-trivial components of ${}^*S_{a}$ are
\begin{eqnarray}
{}^*S^2_{~\theta\phi}=-{}^*S^2_{~\phi\theta} &=& -2 H_0 e^{2H_0t}\,r^2\sin(\theta) ,
\nonumber\\
{}^*S^3_{~r\phi}=-{}^*S^3_{~\phi r}  &=& 2 H_0 e^{2H_0t}\,r\sin(\theta) ,
\nonumber\\
{}^*S^4_{~r\theta}=-{}^*S^4_{~\theta r}&=& -2 H_0 e^{2H_0t}\,r ,
\end{eqnarray}
or the non-dual non-trivial components are
\begin{eqnarray}
S_{212}=-S_{221}&=& -2H_0 ,
\nonumber\\
S_{313}=-S_{331}&=& -2H_0 , 
\nonumber\\
S_{414}=-S_{441}&=& -2H_0 .
\end{eqnarray}

The part of the superpotential that comes into the FEs in terms of coordinates is $S_{(ab)}^{~~\mu}$ and  $S_{[ab]}^{~~\mu}$.
The non trivial components of $S_{[ab]}^{~~\mu}$ are
\begin{eqnarray}
S_{[12]}^{~~r} = - S_{[21]}^{~~r} &=& \frac{H_0}{e^{H_0t}} ,
\nonumber\\
S_{[13]}^{~~\theta} =-S_{[31]}^{~~\theta} &=& \frac{H_0}{e^{H_0t}\,r} ,
\nonumber\\
S_{[14]}^{~~\phi} = -S_{[41]}^{~~\phi}&=& \frac{H_0}{e^{H_0t}\,r\sin(\theta)} .
\end{eqnarray}
The non trivial components of $S_{(ab)}^{~~\mu}$ are
\begin{eqnarray}
S_{(12)}^{~~r}      =S_{(21)}^{~~r} &=& -\frac{H_0}{e^{H_0t}} ,
\nonumber\\
S_{(13)}^{~~\theta} =S_{(31)}^{~~\theta} &=& -\frac{H_0}{e^{H_0t}\,r} ,
\nonumber\\
S_{(14)}^{~~\phi}   =S_{(41)}^{~~\phi} &=& -\frac{H_0}{e^{H_0t}\,r\sin(\theta)} ,
\nonumber\\
S_{22}^{~~t}=S_{33}^{~~t}=S_{44}^{~~t} &=& -2H_0 .
\end{eqnarray}
The torsion scalar is $T=6H_0^{\,2}$, the axial part of the torsion scalar is zero, and the magnitude of the vector part is $-9H_0^2$ ($k=0,   T=6H_0^{\,2}$; for reference, the vector part is $V^2=-9H_0^{\,2}$ and the axial part $A^2=0$).

Assuming a comoving perfect fluid as the matter source, the FEs in the TdS geometry reduce to
\begin{equation}
\kappa\,\rho = -\kappa\,P = -\frac{1}{2}F(T_0) + 6F'(T_0)\,H_0^{\,2} ,
\end{equation}
where necessarily $\rho$ and $P$ are constants.  The equations formally reduce to their GR counterparts only when $F(T)=T$. We note that the effective equation of state
\begin{equation}
\omega_{eff}=\frac{P}{\rho}=-1 ,
\end{equation}
is the same as its GR counterpart; however, the effective cosmological constant $\Lambda_{eff}\equiv\kappa \rho$ depends in principle on  the two parameters $F(T_0)$ and $F'(T_0)$. The energy-momentum conservation eqn. 
\begin{equation}
{\dot{\rho}}  + 3\frac{\dot{a}}{a}(\rho + P) = 0 ,
\end{equation}
is identically satisfied.

\section{Applications}

If $T = \text{const.}$, then the FEs for $F(T)$ teleparallel gravity are equivalent to a rescaled version of TEGR (which looks like GR with a cosmological constant and a rescaled coupling constant) \cite{Krssak:2018ywd}. In the case of TEGR, where $F(T)=T$, eqn. \eqref{temp2} vanishes.  For $F(T) \neq T$, the variation of the gravitational Lagrangian by the flat spin-connection is equivalent to the antisymmetric part of the FEs in eqn. \eqref{temp2} \cite{Krssak:2018ywd}.  In addition, since the canonical energy momentum is symmetric, $\Theta_{[ab]} = 0,$ then the antisymmetric part of the FEs \eqref{temp2} limit the possible solutions of spacetimes.

Cosmological models in flat ($k = 0$) TRW models have been studied in \cite{Bahamonde:2021gfp,Cai_2015} (also see references within). For example, simple power-law scale factor solutions in specific $F(T)$ models (using a priori ansatz such as a polynomial functions) have been investigated. Futhermore, the late time behaviour of the usual de Sitter spacetime, and in particular its stability (as a fixed point), has been studied in \cite{Bahamonde:2021gfp} (see also \cite{Bohmer}).

\subsection{Stability}

Let us simply consider the linear perturbations of the scalar quantities
$T,\,\rho,\,P$ in the TdS solution:
\begin{subequations}
\begin{eqnarray}
T &=& 6\,H_0^{\,2} + T_1,  \label{pert1a} 
\\
\kappa\rho  &=& - \frac{F(T_0)}{2} + 6F'(T_0)\,H_0^{\,2} + \kappa \rho_1, \label{pert1b}
\\
 \kappa P &=&   - \left(-\frac{F(T_0)}{2} + 6F'(T_0)\,H_0^{\,2}\right) + \beta\,\kappa \rho_1, \label{pert1c}
\end{eqnarray}
\end{subequations}
where we have used the zeroth order TdS expressions above with $T_0 \equiv 6H_0^{\,2}$,
and we assume the parameter $\beta \geq -\frac{1}{3}$ from the energy conditions. 

To first order the antisymmetric FEs yield
\begin{equation}
S_{0~[ab]}^{~~\gamma} T_{1,\gamma} = 0 ,
\end{equation}
where the non trivial zeroth order components  of  $S_{[ab]}^{~~\mu}$ are given (by eqns. \eqref{pert1a} to \eqref{pert1c}) above, so that trivially $T_{1,\gamma} = 0$ and hence $T_1 = T_1(t)$.

The FEs and the conservation law then imply that
\begin{subequations}
\begin{eqnarray}
\left[-\frac{1}{2}F'(T_0) + 6F''(T_0) H_0^{\,2}\right]\,T_1 &=& \kappa\rho_1,  \label{pert2a}
\\
- 3H_0(1 + \beta) \rho_1 &=& \dot{\rho_1}. \label{pert2b}
\end{eqnarray}
\end{subequations}
We immediate have that $\rho_1 = \rho_1(t)$, so that from the conservation eqn. $\rho_1 \rightarrow 0$ to the future (since $H_0 > 0$), whence (except in the degenerate case in which the term in square brackets above might be zero), $T_1 \rightarrow 0$, and hence $T \rightarrow T_0$, a constant, so that the resulting geometry is formally equivalent to that of GR and the corresponding stability results follow.

\subsection{The function $F(T)$}

A number of particular examples of $F(T)$ theories studied in the literature include polynomial functions in $T$, and especially quadratic $T^2$ theory 
\cite{Bahamonde:2021gfp,Cai_2015}. 
Recently, it has been shown that the theory { \cite{saridakis} }
\begin{equation}
F(T) = -\Lambda+T+\gamma\,T^2 , \label{QFT}
\end{equation}
can alleviate a variety of cosmological tensions. It is also of interest to study the theory with
\begin{eqnarray}
F(T) = T + \gamma T^{\tilde{\beta}},
\end{eqnarray}
where $\tilde{\beta}>0$ ({ where $\tilde{\beta}$ is potentially} a positive integer greater than 2 for sufficient differentiability), or an exponential function
\begin{equation}
F(T) = F(0)\,\left[\exp\left(2\gamma\,T\right)-1\right]\, \sim \, F(0)\,\left[T + \gamma\,T^2+ ...\right]
\end{equation}
for small $T$.

\section{Analogues of Einstein spaces}

In Riemannian geometries the subclass of Einstein spaces with ${ \tilde{R}_{ab}} = \lambda g_{ab}$,  and which contain the de Sitter geometry as a special case, are of interest \cite{EinsteinSpace}. We note that for  Einstein spaces, in which the algebraic structure of the Ricci tensor ${ \tilde{R}_{ab}}$ is special, the frame components of the Ricci tensor  ${ \tilde{R}_{ab}}$ are constants and its covariant derivative is zero.

It may be of interest to study analogous particular teleparallel geometries with special properties (but non-constant $T$) which include the TdS geometry. Possible examples might include { teleparallel geometries} with special algebraic, topological or symmetry properties. Such examples often lead to $\tilde{R}_{ab} = \lambda g_{ab}$ and hence to a subcase of the usual Einstein spaces (which implies the energy momentum tensor $\Theta_{ab} = \lambda g_{ab}$).

\subsection{Algebraic}

Examples include:
\begin{enumerate}
\item The components of the torsion tensor are constant or its covariant derivative is zero.

\item The algebraic properties of some tensor constructed from the torsion tensor are special.
\end{enumerate}

In addition, the torsion tensor can be invariantly decomposed algebraically into its vectorial part $V_a$ (eqn. \eqref{tensorpartvec}), axial part $A_a$ (eqn. \eqref{tensorpartxi}) and tensorial part $T_{abc}$ (eqn. \eqref{tensorpartsten}). Special invariantly defined subcases occur when the axial and tensorial parts are zero  (or tensorial and vector parts are zero). For example, the TdS geometry has vanishing axial part zero \cite{McNutt_Coley_vdH2022}.

\subsection{Topological}

Special classes of spacetimes have topological properties such as product (or warped product) structures, and hence decomposable and reducible structures \cite{topology}. Unlike the usual Riemannian case, it is the frame and spin-connection (and/or the torsion tensor) that should have these splitting properties (rather than the curvature and metric).

In the Riemannian case, the de Rham decomposition theorem states that if the holonomy group of a simply-connected Riemannian manifold preserves a proper subspace of the tangent space (i.e., is reducible), then the tangent space is decomposable (into holonomy invariant subspaces) and the manifold is (locally) isometric to a product manifold (i.e., a Riemannian manifold is locally a
product of Riemannian manifolds with irreducible holonomy algebras). The Lorentzian case was studied in \cite{BCH}, utilizing Wu’s theorem \cite{wu}, that asserts that every simply-connected, complete semi Riemannian manifold is isometric to a product of simply-connected, complete semi-Riemannian manifolds, of which one can be flat and all others are indecomposable or “weakly-irreducible” (i.e., with no nondegenerate invariant subspace under holonomy representation).

\subsection{Symmetries}

The teleparallel de Sitter  (TdS) geometry has a $G_7$ Lie group of affine symmetries which is the semi direct product of { a} one-dimensional subgroup $O(1,4)$ and { a} six dimensional Euclidean group $E(3)$. We could define an appropriate generalization of TdS by considering spaces with a subalgebra of affine symmetries. For example, if we consider the subalgebra generated by the 3 rotational affine symmetries $X_4 - X_6$ (spherical symmetry) and the special affine symmetry $X_7= -\frac{1}{H_0} \partial_t + r \partial_r$  (eqn. \eqref{X7_desitter}), we obtain a subclass of geometries with { a} 4-dimensional Lie subgroup as its affine symmetry group, which we shall define as Einstein Teleparallel (ET) geometries. The TdS geometry is then a special case in which there are an additional 3 affine symmetries ($X_1 - X_3$) \cite{McNutt_Coley_vdH2022,newColey2023}.

\section{Einstein teleparallel geometries}

Relative to the representation for the isotropy group used in  \cite{McNutt_Coley_vdH2022}, we can determine the most general frame and spin-connection with the three spherically symmetric affine { frame symmetry generators} $X_4 - X_6$ along with the fourth additional affine symmetry $X_7$ defined above (see \cite{McNutt_Coley_vdH2022} and also the case $\lambda=0$ in \cite{newColey2023}). We can also choose a new coordinate system to "diagonalize'' the frame.  The resulting veilbein is:

\begin{equation} h^a_{~\mu} = \left[ \begin{array}{cccc} A_1(t,r) & 0 & 0 & 0 \\ 0 & A_2(t,r) & 0 & 0 \\ 0 & 0 & A_3(t,r) & 0 \\ 
0 & 0 & 0 & A_3(t,r) \sin(\theta) \end{array}\right] . \label{VB:SS} 
\end{equation}
With this choice of invariant symmetry frame, we can now obtain the most general  metric compatible connection \cite{McNutt_Coley_vdH2022}: 
{
\begin{align} \label{Con:SS}
& \omega_{341} = X_1(t,r), \,\quad\quad\quad\quad \omega_{342} = X_2(t,r), 
\nonumber\\ 
& \omega_{233} = \omega_{244} = X_3(t,r),\quad \omega_{234} = -\omega_{243} = X_4(t,r), 
\nonumber\\
& \omega_{121} = X_5(t,r), \,\quad\quad\quad\quad \omega_{122} = X_6(t,r), 
\nonumber\\
& \omega_{133} = \omega_{144} = X_7(t,r), \quad \omega_{134} = -\omega_{143} = X_8(t,r), 
\nonumber\\
& \omega_{344} = - \frac{\cos(\theta)}{A_3 \sin(\theta)}. 
\end{align} 
}

\noindent 
Finally, to determine the most general connection for a teleparallel geometry we must impose the flatness condition. The resulting equations can be solved so that
any spherically symmetric teleparallel geometry is defined by (the three arbitrary functions in the veilbein \eqref{VB:SS} along with) the following spin-connection components \cite{newColey2023}:
\begin{align}\label{SS:TPcon} 
X_1 &= -\frac{\partial_t \chi}{A_1}, \quad\quad\quad\quad\quad X_2 = -\frac{\partial_r \chi}{A_2},
\nonumber\\
X_3 &= \frac{\cosh(\psi)\cos(\chi)}{A_3}, \,\quad X_4 = \frac{\cosh(\psi)\sin(\chi)}{A_3},
\nonumber\\
X_5 &= -\frac{\partial_t \psi}{A_1}, \quad\quad\quad\quad\quad  X_6 = -\frac{\partial_r \psi}{A_2},
\nonumber\\
X_7 &= \frac{\sinh(\psi) \cos(\chi)}{A_3}, \,\quad X_8 = \frac{\sinh(\psi) \sin(\chi)}{A_3},  
\end{align} 
where $\chi$ and $\psi$ are arbitrary functions of the coordinates $t$ and $r$.

Thus, to determine the conditions of the inclusion of the new symmetry generator, we need only examine the conditions 
$\mathcal{L}_{{\bf X}_7} \bh^a  = 0$, from which we obtain:
\begin{equation} 
A_1 (t,r) = f_1(z),\quad\quad  A_2(t,r) = { \frac{f_2(z)}{r} } , \quad\quad A_3(t,r) = f_3(z),
\end{equation}
where $z \equiv r e^{-C_0 t}$ and $C_0 \equiv -H_0$ is a non-zero constant. The Lie derivatives of the metric $g$ and the spin-connection, respectively, yield $\mathcal{L}_{{\bf X}_7}\,g = 0$ and $\mathcal{L}_{{\bf X}_7} \omega^a_{~bc}= 0$, as desired. It therefore follows that the arbitrary functions in the spin-connection's components must be functions of the form:
\begin{equation}
\chi(t,r) = \chi(z),\quad\quad\quad\quad \psi(t,r) = \psi(z).    
\end{equation}

{For general situations in which the parameter $\lambda \neq 0$, we have that the torsion scalar $T=T(z,r)$ will also contain terms studied in \cite{newColey2023} in $r^{\lambda}$ for $T(z,r)=\frac{T_0(z)}{r^{2\lambda}}$. From the FEs we obtain the following equation in term of $r^{-2\lambda}\,F'(T)$:
\begin{eqnarray}
\kappa\,\left(P + \rho \right) &=& 2\,\left[-g_4\,\left[\frac{\left(h_2 +2\,h_3\right)\,T_0'(z)+ \left(2\,m_3\right)\,T_0(z)}{h_4\,T_0'(z)+m_4\,T_0(z)}\right] +\left(g_2+g_3\right)\right]\,r^{-2\lambda}\,F'(T), 
\nonumber\\
 &\equiv& 2\,H(z)\,r^{-2\lambda}\,F'(T).   \label{114}
\end{eqnarray}
We note that in general with $\lambda \neq 0$, $H(z) \neq 0$  \cite{newColey2023}.

For $\lambda = 0$, as considered here, we obtain that $T(z,r)=T(z)$. In general, without substituting in the solutions of the antisymmetric FEs solutions (and the symmetric FEs), we have that
$H(z) \neq 0$. However, we will see below that, using the solution to the antisymmetric FEs, we find that $H(z)=0$. This then implies that $\rho + P= 0$ . This is consistent with the fact that this is intended as an anologue of an Einstein space.

\subsection{Antisymmetric FEs and solutions}

For $\lambda = 0$, the antisymmetric FEs become (when the torsion scalar $T(z)\neq \text{const}$ and $F(T)$ is not linear in $T$):
\begin{subequations}
\begin{align}
& 0 = \sin\,\chi(z)\,\left[C_0\,f_2(z)\,\sinh\psi(z) + f_1(z)\,\cosh\psi(z)\right] , \label{501a}
\\
& 0 = \cos\,\chi(z)\,\left[C_0\,f_2(z)\,\cosh\psi(z) + f_1(z)\,\sinh\psi(z)\right].   \label{501b}
\end{align}
\end{subequations}
There are two solutions: 
\begin{enumerate}

\item[A:] $\cos\,\chi(z)=0$: $\sin\,\chi(z)=\delta=\pm 1$, $\chi = \left(k+\frac{1}{2}\right)\pi$ and $\tanh \psi(z)=-\frac{f_1(z)}{C_0\,f_2(z)}$.

\item[B:]  $\sin\,\chi(z)=0$: $\cos\,\chi(z)=\delta=\pm 1$, $\chi = k\pi$ and $\tanh \psi(z)=-\frac{C_0\,f_2(z)}{f_1(z)}$.
\end{enumerate}

\subsection{General form of symmetric FEs }

The general symmetric FEs can be expressed in the following form:
\begin{subequations}
\begin{eqnarray}
k_1\,\partial_z\left(\ln(F'(T))\right) &=& g_1  ,\label{504a}
\\
\kappa\,P -\frac{F(T)}{2} &=& 2\,F'(T)\,\left[-k_2\,\partial_z\left(\ln(F'(T))\right) +g_2\right] ,  \label{504b}
\\
\kappa\,\rho +\frac{F(T)}{2} &=& 2\,F'(T)\,\left[-k_3\,\partial_z\left(\ln(F'(T))\right) +g_3\right] , \label{504c}
\\
k_4\,\partial_z\left(\ln(F'(T))\right) &=& g_4 , \label{504d}
\end{eqnarray}
\end{subequations}
where the functions $g_i$ and $k_i$ ($i=1-4$) are explicitly displayed below for solution A (the functions for solution B are displayed in the Appendix). By combining the pair of eqns. \eqref{504b} and \eqref{504c} and then eqns. \eqref{504a} and \eqref{504d}, respectively, we obtain:
\begin{subequations}
\begin{eqnarray}
\partial_z\left(\ln(F'(T))\right) &=& \frac{g_1}{k_1} = \frac{g_4}{k_4} ,   \label{540a}
\\
\kappa\,\left(P+\rho\right) &=& 2\,F'(T)\left[-\left(k_2+k_3\right)\,\partial_z\left(\ln(F'(T))\right) + \left(g_2+g_3\right)\right]\equiv F'(T) H(z).
 \label{540b}
\end{eqnarray}
\end{subequations}

\noindent The general torsion scalar expression respecting eqns. \eqref{540a} and \eqref{540b} is the following:
\small
\begin{align}\label{502}
T(z) =& \frac{1}{f_1^2\,f_2^2\,f_3^2}\Bigg[-4\,f_1\,f_2\,z\,\cos\chi\Bigg(\left[\left(f_3\,f_1\right)'+f_3\,f_2\,\psi'\right]\cosh\psi+\sinh\psi\left[\psi'\,f_1\,f_3+C_0\left(f_2\,f_3\right)'\right]\Bigg)
\nonumber\\
&\quad +4\,z\,\sin\chi\,\cosh\psi\,\chi'\,f_1^2\,f_2\,f_3+4\,C_0\,z\,\sin\chi\,\sinh\psi\,\chi'\,f_1\,f_2^2\,f_3-2\,f_1^2\left(z^2\,f_3'^2+f_2^2\right) 
\nonumber\\
&\quad -4\,z^2\,f_1\,f_3\,f_1'\,f_3'
+4\,C_0^2\,z^2\,f_2\,f_3'\left(f_2'\,f_3+\frac{f_2\,f_3'}{2}\right)\Bigg] .
\end{align}
\normalsize

\subsection*{Coordinates}

There is a class of transformations that maintain the ``diagonal" form of the frame in (\ref{VB:SS}). { However}, in general, this will also lead to a change in the form of the connection and the FEs. Therefore, in order to maintain the comoving nature of the perfect fluid source, we restrict ourselve to transformations of the form $t \rightarrow f(t)$ and $r \rightarrow g(r)$ (i.e., redefining the $r$ and $t$ coordinates). This may simplify the eqns. further. { Even then,} the explicit forms for the symmetry vectors and the similarity variable $z$ (for example) does change. We shall not change coordinates here.

However, since the static case corresponds formally to $C_0=0$, we assume explicitly that $C_0^2 \neq 0$ here. In this case we can then effect a simple time change to set $C_0^2 =1$, which we shall do hereafter. This has the effect of changing the metric, $-dt^2 \rightarrow  - \frac{1}{H_0^2}dt^2$. In the exact TdS solutions $C_0^2 =1$ and the metric changes accordingly.

\subsection{Solution A}

The symmetric FE components (see eqns. \eqref{504a} to \eqref{504d}) for solution A (with $C_0^2 = 1$) are explicitly given by (the symmetric FEs components for solution B are displayed in the Appendix):
\small
\begin{subequations}
\begin{align}
g_1 =& \Bigg[-z^2\,f_1\,f_2\,f_3\,f_3''\left(f_1^2+f_2^2\right)-z^2\,f_1^2\,f_2\,f_3^2\,f_1''+z^2\,f_1\,f_2^2\,f_3^2\,f_2''+z^2\,f_3'^2\left(-f_1\,f_2^3 + f_1^3\,f_2\right)  \label{511aa}
\nonumber\\
&+f_3\,f_3'\,\left(f_2^2+f_1^2\right)\left(z^2\,f_2\,f_1'+f_1\,\left(z^2\,f_2'-z\,f_2\right)\right)+f_3^2\,f_1'\,\left(\left(f_1^2-f_2^2\right)\,z^2\,f_2'-z\,f_1^2\,f_2\right)
\nonumber\\
&+f_1\,f_2^2\,\left(z\,f_3^2\,f_2'-f_1^2\,f_2\right)\Bigg]  ,
\\
g_2 =& \frac{1}{f_1^3\,f_2^2\,f_3^2}\Bigg[-z^2\,f_1\,f_2^2\,f_3\,f_3''+z^2\,f_1\,f_3'^2\,\left(f_1^2-f_2^2\right)+f_3\,f_3'\Bigg(z^2\,f_1'\left(f_2^2+2\,f_1^2\right)
\nonumber\\
&-f_1\,f_2\,\left(z^2\,f_2'+z\,f_2\right)\Bigg)\Bigg] , \label{511ab}
\\
g_3 =& \frac{1}{f_1^2\,f_2^3\,f_3^2}\Bigg[-z^2\,f_3''\,f_3\,f_1^2\,f_2+f_2\,f_3'^2\,z^2\left(f_2^2-f_1^2\right)-f_3'\Bigg(-z^2\,f_3\,f_2'\left(2\,f_2^2+f_1^2\right)
\nonumber\\
&+f_1\,f_2\left(z^2\,f_3\,f_1'+zf_1\,f_3\right)\Bigg)\Bigg] , \label{511ac}
\\
g_4 =& \Bigg[\ln \left(\frac{f_1\,f_2}{z\,f_3'}\right)\Bigg]' , \label{511ad}
\end{align}
\end{subequations}

\begin{subequations}
\begin{align}
k_1 =& \left(z^2\,f_1\,f_2\,f_3\right)\,\Bigg[f_3'\left(f_1^2+f_2^2\right)+f_3\,\left(f_1\,f_1'-f_2\,f_2'\right)\Bigg] , \label{511ba}
\\
k_2 =& \frac{z^2\,f_3'}{f_1^2\,f_3} , \label{511bb}
\\
k_3 =& \frac{z^2\,f_3'}{f_2^2\,f_3}  , \label{511bc}
\\
k_4 =& 1 . \label{511bd}
\end{align}
\end{subequations}
\normalsize

We simplify the symmetric FEs for solution A into eqn. \eqref{540a}, to obtain:
\begin{align}\label{513}
& \Bigg[-z^2\,f_1\,f_2\,f_3\,f_3''\left(f_1^2+f_2^2\right)-z^2\,f_1^2\,f_2\,f_3^2\,f_1''+z^2\,f_1\,f_2^2\,f_3^2\,f_2''+z^2\,f_3'^2\left(-f_1\,f_2^3 +f_1^3\,f_2\right)
\nonumber\\
&+f_3\,f_3'\,\left(f_2^2+f_1^2\right)\left(z^2\,f_2\,f_1'+f_1\,\left(z^2\,f_2'-z\,f_2\right)\right)+f_3^2\,f_1'\,\left(\left(f_1^2-f_2^2\right)\,z^2\,f_2'-z\,f_1^2\,f_2\right)
\nonumber\\
&+f_1\,f_2^2\,\left(z\,f_3^2\,f_2'-f_1^2\,f_2\right)\Bigg]\Bigg[\left(z^2\,f_1\,f_2\,f_3\right)\,\Bigg[f_3'\left(f_1^2+f_2^2\right)+f_3\,\left(f_1\,f_1'-f_2\,f_2'\right)\Bigg]\Bigg]^{-1}
\nonumber\\
&=-\left(\ln\left(\frac{z\,f_3'}{f_1\,f_2}\right)\right)' .
\end{align}

Using the eqns. above and the symmetric FE, we find that the right-hand side of eqn. \eqref{540b} vanishes, thus giving $P+\rho = 0$. So, for $\lambda=0$, we only have $\kappa P(t,r)=- \kappa \rho(t,r) \equiv -\Lambda_0$, regardless of the form of $F(T(z))$. This is consistent with the fact that this is intended as an anologue of an Einstein space. This means that we have the following relations:
\begin{eqnarray}\label{522}
\left(k_2+k_3\right)\,g_4 &=& \left(g_2+g_3\right) ,
\nonumber\\
-\left(\ln\left(\frac{z\,f_3'}{f_1\,f_2}\right)\right)' &=& g_4 =\frac{g_2+g_3}{k_2+k_3} .
\end{eqnarray}
By putting the eqns. (\ref{513}) and (\ref{522}) together, we arrive at the following superrelation:
\begin{eqnarray}\label{523}
-\left(\ln\left(\frac{z\,f_3'}{f_1\,f_2}\right)\right)' = g_4 =\frac{g_2+g_3}{k_2+k_3}=\frac{g_1}{k_1}.
\end{eqnarray}
The torsion scalar expression for solution A simplifies and is given explicitly by:{
\small
\begin{align}\label{516}
T(z) =& \frac{2\,z^2\,f_3'^2(z)\left(f_2^2(z)-f_1^2(z)\right)+4z^2\,f_3(z)\,f_3'(z)\left(f_2(z)\,f_2'(z)-f_1(z)\,f_1'(z)\right)-2\,f_1^2(z)\,f_2^2(z)}{f_1^2(z)\,f_2^2(z)\,f_3^2(z)} .
\end{align}
\normalsize
}

\subsection{Explicit equations: Summary}

First we satisfied the Antisymmetric FEs (eqns. \eqref{501a} and \eqref{501b}) leading to solutions A and B. Then we set $C_0^2 =1$ in all of our physical quantities. For the symmetric FEs, we have eqns. \eqref{504a} to \eqref{504d} with $g_i$ and $k_i$  ($i=1-4$) given by eqns. \eqref{511aa} to \eqref{511bd}, respectively. We restrict to solution A hereafter, whence eqn. \eqref{540a} becomes eqn. \eqref{513} (see also eqns. \eqref{522} and \eqref{523}). After finding $H(z)=0$, eqn. \eqref{540b} implies that $\rho+P=0$. Then the torsion scalar $T(z)$ is given explicitly by eqn. \eqref{516}. The FE described by eqn. \eqref{504d}, using $k_4=1$ and the expression above for $g_4$, is replaced by the {\em{first integral}}
\begin{equation}\label{FI}
\frac{dF}{dT}=B_0\left(\frac{f_1\,f_2}{z\,f_3'}\right) ,
\end{equation}
which is a function of $z$, and where $B_0$ is a constant. The remaining FE described by eqn. \eqref{504c} yields an eqn. for $F(T(z))$ in terms of $z$. We have that $\kappa \rho=- \kappa P=\Lambda_0$, where we have used solution A and set $C_0^2 =1$. In principle, the three remaining FEs, eqns. \eqref{504c}, \eqref{513} and \eqref{FI} are then $3$ ODEs for the 3 functions $f_i(z)$ for a given $F(T(z))$.

For example, for the quadratic function $F(T)$ explicitly given by eqn. \eqref{QFT}, we find that $F^\prime = 1 + 2 \gamma T, F^{\prime\prime} = 2 \gamma$ which, when substituted into eqns.
\eqref{FI}, yields 
\begin{equation} 
T =   -\frac{1}{2 \gamma} + \frac{B_0}{2 \gamma} \left(\frac{f_1\,f_2}{z\,f_3'}\right),
\end{equation}
whence using \eqref{516} yields a first order ODE for the $f_i$.  Eqn. \eqref{504c} then yields a quadratic eqn.  for $T$ in terms of functions of $z$ (including a term $\frac{dT}{dz}$) which { upon} using eqn. \eqref{516} yields a second order ODE for the $f_i$. Eqn. \eqref{513} constitutes another second order ODE for the $f_i$.


We note again that all potential solutions considered here (other than TdS explicitly) are not analogues of GR solutions and are consequently new.

\section{Power Law Solutions}

To find solutions to the symmetric part of the FEs describing an Einstein teleparallel geometry,
we use the following ansatz:
\begin{eqnarray}\label{506}
f_1(z)=a_0\,z^a , \quad\quad f_2(z)=b_0\,z^b , \quad\quad f_3(z)=c_0\,z^c,
\end{eqnarray}
where $a_0,\,b_0,\,c_0 \neq 0$. Using the { solution A}, from eqn. \eqref{504c} we obtain:
\begin{align}\label{507}
\Lambda_0=& -\frac{F(T(z))}{2}+2\,F'(T(z))\,\Bigg[\frac{c\,(2b+c)}{a_0^2}\,z^{-2a}-\frac{c\,(2a+c)}{b_0^2}\,z^{-2b}\Bigg] .
\end{align}
We need the exact expressions for $T(z)$ and $F(T(z))$ before simplifying further. Using $g_1=g_4\,k_1$, we obtain from eqn. \eqref{513}:
\begin{align}\label{508}
0=& z^{a+b}\Bigg[a_0^2\,c_0^2\,\left(c^2-2a^2+ac\right)\,z^{2a+2c}-b_0^2\,c_0^2\,\left(c^2-2b^2+bc\right)\,z^{2b+2c}-a_0^2\,b_0^2\,z^{2a+2b}\Bigg] .
\end{align}

\noindent For the torsion scalar, we obtain:
\begin{align}\label{509}
T(z) =& 2\Bigg[\frac{c(2b+c)}{a_0^2}\,z^{-2a}-\frac{c(2a+c)}{b_0^2}\,z^{-2b}-\frac{1}{c_0^2}z^{-2c}\Bigg] .
\end{align}
The derivative $T'(z)$ is:
\begin{align}\label{510}
T'(z) =& -4\,c\left[\frac{(2b+c)\,a}{a_0^2}\,z^{-2a-1}-\frac{(2a+c)\,b}{b_0^2}\,z^{-2b-1}-\frac{1}{c_0^2}z^{-2c-1}\right] .
\end{align}

\noindent From eqn. \eqref{FI} we obtain:
\begin{eqnarray}\label{513a}
\frac{dF(T)}{dT} &=& \frac{{ B_1}}{c}\,z^{a+b-c} ,
\end{eqnarray}
where ${ B_1}$ is a constant (the degenerate case $c=0$ is not included). For $c=a+b$ we get exactly the TEGR solution. We will consider $a+b \neq c$ for a non-trivial $F(T)$ solution.

We multiply by the derivative of the torsion scalar to obtain:
\begin{eqnarray}\label{514}
\frac{dF(T(z))}{dz} &=&\frac{dF(T)}{dT}\,T'(z)  = -4{ B_1}\,\left[\frac{(c+2b)\,a}{a_0^2}\,z^{b-c-a-1}-\frac{(c+2a)\,b}{b_0^2}\,z^{a-c-b-1}-\frac{1}{c_0^2}z^{a+b-3c-1}\right] .
\nonumber\\
\end{eqnarray}
From this equation, we may integrate $\frac{dF(T(z))}{dz}$ w.r.t $z$ (depending on the power of $z$ - some situations may lead to $z^{-1}$ and care should be taken). From eqns. \eqref{508}, we can solve the symmetric FEs for $a$, $b$ and $c$ and then verify eqn. \eqref{507} by taking into account eqns. \eqref{509} to \eqref{514} for each possible situation. In principle, there may be some free parameters remaining.

\subsection{The possible solutions}

If we consider eqn. \eqref{508} only there are, in principle, the following possibilities:
\begin{enumerate}

\item $a=b$ (with $a_0^2\,b_0^2 \neq 0$), which is not possible because we need to satisfy:
\begin{align}\label{520}
0= \left(c^2-2a^2+ac\right)\,c_0^2\Bigg[a_0^2-b_0^2\Bigg] \quad\text{and}\quad  0= a_0^2\,b_0^2  .
\end{align}
This case includes the special case $a=b=c$.

\item  $a=c$: which leads to:
\begin{align}\label{521}
0=& -\Bigg[c_0^2\,\left(c^2-2b^2+bc\right)+a_0^2\Bigg]\,b_0^2 ,
\end{align}
which gives the following relation between $a$ and $b$:
\begin{eqnarray}\label{521b}
(a+2b)(a-b)=-\frac{a_0^2}{c_0^2},
\end{eqnarray}
where, according to eqn. \eqref{521b}, $-2b < a < +b$ for viable solutions.

\item $b=c$: which leads to
\begin{align}
0=& \Bigg[c_0^2\,\left(b^2-2a^2+ab\right)-b_0^2\Bigg]. \label{521c}
\end{align}
Eqn. \eqref{521c} leads to the following relation between $a$ and $b$:
\begin{eqnarray}\label{521d}
(b+2a)(b-a)=\frac{b_0^2}{c_0^2},
\end{eqnarray} 
where it follows that $-\frac{b}{2} < a < +b$ for viable solutions.

\end{enumerate}

\subsection{Case I: $a=c$}

For $a=c$ and $\frac{a_0^2}{c_0^2}=-(a+2b)(a-b)$ (eqn. \eqref{521b}), we then compute eqns. \eqref{509}, \eqref{513a} and \eqref{514}:
\begin{subequations}
\begin{align}
T(z) =& 2\Bigg[(a+2b)(2a-b)\,\frac{z^{-2a}}{a_0^2}-\frac{3a^2}{b_0^2}\,z^{-2b}\Bigg] , \label{530a}
\\
\frac{dF(T)}{dT}=& F'(T(z)) = \frac{{ B_1}}{a}\,z^{b} , \label{530b}
\\
\frac{dF(T(z))}{dz} =& -4{ B_1}\,\left[\frac{(a+2b)(2a-b)}{a_0^2}\,z^{b-2a-1}-\frac{3a\,b}{b_0^2}\,z^{-b-1}\right]  . \label{530c}
\end{align}
\end{subequations}
When we attempt to integrate eqn. \eqref{530c}, we see that there may be special cases ($b=0$ or $a=\frac{b}{2}$) which yield a $z^{-1}$ power term, but in each case the corresponding coefficient is zero and therefore no power of $z^{-1}$  terms ever occurs in eqn. \eqref{530c}. From this, we can integrate eqn. \eqref{530c} giving the relation:
\begin{eqnarray}\label{531a}
F(T(z)) =& -4{ B_1}\,\left[-\frac{(a+2b)}{a_0^2}\,z^{b-2a}+\frac{3a}{b_0^2}\,z^{-b}\right]  +C ,
\end{eqnarray}
where $C$ is a constant. By substituting eqn. \eqref{531a} into eqn. \eqref{507}, we find that:
\begin{align}\label{532a}
\Lambda_0 +\frac{C}{2} =&  2{ B_1}\,\left[\frac{(a+2b)}{a_0^2}-\frac{(a+2b)}{a_0^2}\right]\,z^{b-2a}=0.
\end{align}
This means that we have $C=-2\Lambda_0$ in eqn. \eqref{531a} ($b \neq 0$). This constitutes a one parameter ($b$ say) family of solutions. It is possible to locally find the inverse $z=z(T)$ from eqn. \eqref{530a} to explicitly find $F(T)$.

As a simple albeit special example, if $a=\frac{b}{2}$, eqns. \eqref{530a} and \eqref{530b} become:
\begin{subequations}
\begin{align}
z^{b} =& \sqrt{\frac{3}{2}}\,\frac{b}{b_0\,\sqrt{-T}} , \label{533a}
\\
\frac{dF(T)}{dT}=& \frac{\sqrt{24}\,{ B_1}}{2\,b_0\,\sqrt{-T}}\equiv -\frac{F_1}{2\sqrt{-T}} , \label{533b}
\end{align}
\end{subequations}
where $F_1$ is a constant. We integrate eqn. \eqref{533b} w.r.t. to $T$ to obtain:
\begin{eqnarray}\label{534}
F(T) =& F_1\,\sqrt{-T}+C ,
\end{eqnarray}
for $T \leq 0$. By substituting eqn. \eqref{534} into eqn. \eqref{507}, we get that:
\begin{align}\label{535}
C=& - \frac{5\,b\,b_0\, F_1}{2\,a_0^2}\,\sqrt{\frac{2}{3}}-2\Lambda_0 \equiv F_0-2\Lambda_0,
\end{align}
where $F_0$ is also a constant.

\subsection{Case II: $b=c$}

For $b=c$ and $\frac{b_0^2}{c_0^2}=(b+2a)(b-a)$ (eqn. \eqref{521d}), we must evaluate eqns. \eqref{509}, \eqref{513a} and \eqref{514} (for $a \neq 0$):
\begin{subequations}
\begin{align}
T(z) =& 2\Bigg[\frac{3b^2}{a_0^2}\,z^{-2a}-\frac{(2b-a)(2a+b)}{b_0^2}\,z^{-2b}\Bigg]  ,\label{544a} 
\\
\frac{dF(T)}{dT}=& F'(T(z)) = \frac{{ B_1}}{b}\,z^{a} ,\label{544b} 
\\
\frac{dF(T(z))}{dz} =& -4{ B_1}\,\left[\frac{3b\,a}{a_0^2}\,z^{-a-1}-\frac{(2b-a)(2a+b)}{b_0^2}\,z^{a-2b-1}\right]  .
  \label{544c}
\end{align}
\end{subequations}
When we attempt to integrate eqn. \eqref{544c}, we see that there may be special cases when $a=0$ or $a=2b$ which yield $z^{-1}$ power terms, but in each case the corresponding coefficient is zero and therefore such a power of $z^{-1}$ never occurs in eqn. \eqref{544c}. We can consequently integrate eqn. \eqref{544c}, giving the relation:
\begin{eqnarray}\label{541}
F(T(z)) =& 4{ B_1}\,\left[\frac{3b}{a_0^2}\,z^{-a}-\frac{(2a+b)}{b_0^2}\,z^{a-2b}\right]+C .
\end{eqnarray}
By substituting eqn. \eqref{541} into eqn. \eqref{507}, we find that:
\begin{align}\label{542}
\Lambda_0 +\frac{C}{2} =&  -2{ B_1}\,\left[\frac{(2a+b)}{b_0^2}-\frac{(2a+b)}{b_0^2}\right]\,z^{a-2b} = 0.
\end{align}
This again implies that $C=-2\Lambda_0$ as an integrating constant in eqn. \eqref{541}. In general, it is possible to locally find the inverse $z=z(T)$ from eqn. \eqref{544a} then use it to compute $F(T)$. Note that from eqns. \eqref{544a} and \eqref{541}, in the TEGR case we must have $a=0$ (which corresponds to the special case $a+b=c$), where
\begin{align}
z^{-2b} =& \frac{3b_0^2}{2\,a_0^2}-\frac{b_0^2}{4b^2}\,T. \label{543a}
\end{align}

\section{Concluding remarks}

Since the de Sitter geometry (and Einstein spaces in more generality) within GR have a number of important mathematical and physical applications, we have studied TdS geometries and their generalizations in theories  of gravity based on teleparallel geometries. In particular, we have investigated a class of Einstein teleparallel geometries which have a $4$-dimensional Lie algebra of affine symmetries. We solved the resulting antisymmetric FEs and we displayed all of the remaining governing eqns. explictly. We then investigated power law solutions and displayed two  one-parameter families (within solution class A) of explicit Einstein teleparallel solutions (where we can implicitly find the form of $F(T)$ for each parameter value). Very few explicit non-trivial exact solutions are known within teleparallel gravity, and the  two one-parameter families of exact power law solutions obtained (being possible analogues of Einstein spaces) may have a number of important applications in cosmology and astrophysics.


\section*{Acknowledgments}

\noindent AAC and RJvdH are supported by the Natural Sciences and Engineering Research Council of Canada. RJvdH is also supported by the W.F. James Chair of Studies in the Pure and Applied Sciences at St.F.X. AL is supported by an AARMS fellowship. DDMcN is supported by the Norwegian Financial Mechanism 2014-2021 (project registration number 2019/34/H/ST1/00636).



\appendix

\section{Expressions for solution B}\label{appena}

The functions in the symmetric FEs  (eqns. \eqref{504a} to \eqref{504d}) { for} solution B are explicitly given by:
\small
\begin{align}\label{503a}
g_1 =& \Bigg[-z^2\,f_1\,f_2\,f_3\,f_3''\,\left(f_1^2+f_2^2\right)-z^2\,f_1^2\,f_2\,f_3^2\,f_1''+ \,z^2\,f_1\,f_2^2\,f_3^2\,f_2''+z^2\,f_3'^2\,\left(-f_1\,f_2^3+f_1^3\,f_2\right)
\nonumber\\
& + f_3\,f_3'\,\left(f_1^2+ \,f_2^2\right)\left(z^2\,\left(f_1\,f_2\right)'-z\,f_1\,f_2\right)+f_3^2\,f_1'\,\left(z^2\,f_2'\left(f_1^2-f_2^2\right)-z\,f_1^2\,f_2\right)
\nonumber\\
&+f_1\,f_2^2\,\left(z\,f_3^2\,f_2'-f_1^2\,f_2\right)\Bigg] ,
\nonumber\\
g_2 =& \frac{1}{f_1^3\,f_2^2\,f_3^2}\Bigg[-z^2\,f_1\,f_2^2\,f_3\,f_3''+z^2\,f_3'^2\left(-f_1\,f_2^2+f_1^3\right)+2\,f_3'{\Bigg(}z^2\,f_3\,f_1'\left(\frac{f_2^2}{2}+f_1^2\right)
\nonumber\\
&-\frac{f_1\,f_2}{2}\Bigg[z^2\,f_3\,f_2'-\delta\,z\Bigg(\frac{f_1^2-f_2^2}{\sqrt{1-\frac{f_2^2}{f_1^2}}}\Bigg)+z\,f_2\,f_3\Bigg]{\Bigg)}+\delta\,z\,f_1\,f_2\,f_3\,\frac{f_1\,f_1'-f_2\,f_2'}{\sqrt{1-\frac{f_2^2}{f_1^2}}}\Bigg] ,
\nonumber\\
g_3 =& \frac{1}{f_1^2\,f_2^3\,f_3^2}\Bigg[-z^2\,f_1^2\,f_2\,f_3\,f_3''+f_3'^2\,z^2\left(f_2^3-f_1^2\,f_2\right)-f_3'{\Bigg[}-z^2\,f_3\,f_2'\left(2\,f_2^2+f_1^2\right)
\nonumber\\
&+f_1\,f_2\,{\Bigg(z^2\,f_1'\,f_3+\Bigg[\delta\,z\,f_2\,\Bigg(\frac{f_1^2-f_2^2}{f_1\,\sqrt{1-\frac{f_2^2}{f_1^2}}}\Bigg)+z\,f_1\,f_3\Bigg]\Bigg)\Bigg]}-\delta\,z\,f_1\,f_2^2\,f_3\Bigg(\frac{f_1\,f_1'-f_2\,f_2'}{f_1\,\sqrt{1-\frac{f_2^2}{f_1^2}}}\Bigg)\Bigg] ,
\nonumber\\
g_4 =& \Bigg[\ln \left(\frac{f_1\,f_2}{z\,f_3'}\right)\Bigg]' ,
\end{align}
\begin{eqnarray} \label{503b}
k_1 &=& \left[z^2\,f_1\,f_2\,f_3\,f_3'\,\left(f_1^2+f_2^2\right)+f_1\,f_2\,f_3\left(z^2\,f_1\,f_3\,f_1'-f_2\,\left(z^2\,f_3\,f_2'-\delta\,z\,\frac{f_1^2-f_2^2}{\sqrt{1-\frac{f_2^2}{f_1^2}}}\right)\right)\right] ,
\nonumber\\
k_2 &=& \frac{z}{f_1^2\,f_3}\,\left[\frac{\delta\,f_2}{\sqrt{1-\frac{f_2^2}{f_1^2}}}+z\,f_3'\right] ,
\nonumber\\
k_3 &=& \frac{z}{f_2^2\,f_3}\,\left[\frac{\delta\,f_2}{\sqrt{1-\frac{f_2^2}{f_1^2}}}+z\,f_3'\right] ,
\nonumber\\
k_4 &=& \frac{1}{z\,f_3'} \left[\frac{\delta\,f_2}{\sqrt{1-\frac{f_2^2}{f_1^2}}} + z\,f_3'\right] .
\end{eqnarray}
\normalsize
From eqn. \eqref{540a} we explictly get:
\small
\begin{align}\label{505}
&{ \Bigg[z^2\,f_1\,f_2\,f_3\,f_3'\,\left(f_1^2+f_2^2\right)+f_1\,f_2\,f_3\Bigg(z^2\,f_1\,f_3\,f_1'-f_2\,\Bigg[z^2\,f_3\,f_2'-\delta\,z\,\frac{f_1^2-f_2^2}{\sqrt{1-\frac{f_2^2}{f_1^2}}}\Bigg]\Bigg)\Bigg]^{-1}}
\nonumber\\
&\times \Bigg[-z^2\,f_1\,f_2\,f_3\,f_3''\,\left(f_1^2+f_2^2\right)-z^2\,f_1^2\,f_2\,f_3^2\,f_1''+z^2\,f_1\,f_2^2\,f_3^2\,f_2''+z^2\,f_3'^2\,\left(-f_1\,f_2^3+f_1^3\,f_2\right)
\nonumber\\
&+ f_3\,f_3'\,\left(f_1^2+f_2^2\right)\left(z^2\,\left(f_1\,f_2\right)'-z\,f_1\,f_2\right)+f_3^2\,f_1'\,\left(z^2\,f_2'\left(f_1^2-f_2^2\right)-z\,f_1^2\,f_2\right)
\nonumber\\
& +f_1\,f_2^2\,\left(z\,f_3^2\,f_2'-f_1^2\,f_2\right)\Bigg]
\nonumber\\
&= \Bigg[1+\frac{\delta\,f_2}{z\,f_3'\,\sqrt{1-\frac{f_2^2}{f_1^2}}}\Bigg]^{-1}\,\Bigg[\ln \left(\frac{f_1\,f_2}{z\,f_3'}\right)\Bigg]' .
\end{align}
\normalsize

Using eqns. \eqref{503a} and \eqref{503b} and the symmetric FE, we find that the right-hand side of eqn. \eqref{540b} vanishes, thus giving $P+\rho = 0$. So, for $\lambda=0$, we have $P(t,r)=-\rho(t,r) \equiv -\Lambda_0$, regardless of the form of $F(T(z))$. This is consistent with the fact that this is intended as an anologue of an Einstein space. This means that we have the following relation:
\begin{eqnarray}\label{531}
\Bigg[1+\frac{\delta\,f_2}{z\,f_3'\,\sqrt{1-\frac{f_2^2}{f_1^2}}}\Bigg]^{-1}\,\Bigg[\ln \left(\frac{f_1\,f_2}{z\,f_3'}\right)\Bigg]' = \frac{g_4}{k_4} &=&\frac{g_2+g_3}{k_2+k_3} .
\end{eqnarray}
By putting the eqns. \eqref{505} and \eqref{531} together, we arrive at the following superrelation:
\begin{eqnarray}\label{532}
\Bigg[1+\frac{\delta\,f_2}{z\,f_3'\,\sqrt{1-\frac{f_2^2}{f_1^2}}}\Bigg]^{-1}\,\Bigg[\ln \left(\frac{f_1\,f_2}{z\,f_3'}\right)\Bigg]' = \frac{g_4}{k_4} =\frac{g_2+g_3}{k_2+k_3}=\frac{g_1}{k_1}.
\end{eqnarray}

\noindent The expression for the torsion scalar in the second solution then simplifies further.

\end{document}